\documentstyle[12pt,aaspp]{article}
\begin{document}
\title{The Speed of Cooling Fronts and the Functional Form of
the Dimensionless Viscosity in Accretion Disks}
\author{Ethan T. Vishniac and J. Craig Wheeler}
\affil{Department of Astronomy, University of Texas, Austin TX
78712; ethan or wheel@astro.as.utexas.edu}
\begin{abstract}
We examine the speed of inward traveling cooling fronts in accretion
disks.  We show that their speed is determined by the rarefaction
wave that precedes them and is approximately $\alpha_F c_{F} (H/r)^q$,
where $\alpha_F$ is the dimensionless viscosity, $c_{F}$ is the sound speed,
$r$ is the radial coordinate, $H$ is the disk thickness,
and all quantities are evaluated at the cooling front.  The
scaling exponent $q$ lies in the interval $[0,1]$, depending on the
slope of the $(T,\Sigma)$ relation in the hot state. For a Kramer's law
opacity and $\alpha\propto (H/r)^n$, where $n$ is of order unity,
we find that $q\sim 1/2$.  This supports the numerical work of
Cannizzo, Chen and Livio (1995) and their conclusion that $n\approx3/2$ is
necessary to reproduce the exponential decay of luminosity in black hole
X-ray binary systems.  Our results are insensitive to the structure of the
disk outside of the radius where rapid cooling sets in.  In particular,
the width of the rapid cooling zone is a consequence of the cooling
front speed rather than its cause.  We conclude that
the exponential luminosity decay of cooling disks is probably compatible
with the wave-driven dynamo model.  It is not compatible with models with
separate, constant values of $\alpha$ for the hot and cold states.
\end{abstract}

\section{Introduction}

The rate of mass transfer in accretion disks depends on the rate
at which angular momentum can be transferred outward.  This is
normally expressed in terms of a dimensionless viscosity $\alpha$,
which is defined as
\begin{equation}
\alpha\equiv {\nu\over c_s H},
\end{equation}
where $H$ is the disk half-thickness, $c_s$ is the local sound speed,
and $\nu$ is the local effective viscosity (\cite{SS73}).
Initially $\alpha$ was assumed to be constant.
There are now strong grounds, both empirical and theoretical,
for concluding that $\alpha$ must be a variable.
The task is
to combine empirical evidence with theoretical guidance
to construct a self-consistent theory of angular momentum
transport in accretion disks that accounts for the wealth
of observations.  A successful theory is likely to reveal
that angular momentum transport is non-local, so that the concept
of a local viscosity is itself of limited value.

Communication between observational
studies of accretion disks and theories of angular momentum
transport is facilitated by models of time-dependent
accretion disks.  The time dependence is
critical since the emissivity of steady-state disks is
independent of the viscosity.
An excellent review of the history of this
area is given by Cannizzo (1993a; see also Cannizzo 1993b).

The first major constraint on $\alpha$ came from comparison
of limit cycle disk instability models with observations
of dwarf novae.  The models provide a very credible
basic interpretation of the dwarf nova phenomenon, but
only if $\alpha$ is not a constant (Smak 1984).
This constraint does not determine the functional form
of $\alpha$.  Models in which $\alpha$ has one, radially constant
value, $\alpha_{hot}$ in outburst and another lower, but
also radially constant value $\alpha_{cold}$ in quiescence
work about as well as a model in which $\alpha=\alpha_0(H/r)^{n}$
which would apply if $\alpha$ were a function of the sound speed
and hence the temperature of the disk.

Another perspective on the behavior of $\alpha$ can
be obtained by comparing the dwarf novae with
soft X-ray transients.  In the latter case,
the only quantitative work has been done on those
that are black hole candidates, but those are especially
interesting laboratories because of the suspicion that
the compact star lacks a hard surface and an associated
magnetosphere and boundary layer.  Several of the black
hole candidates have outbursts with rapid rise and subsequent
slower decline that are in reasonable agreement with the
same limit cycle disk instability models that account
for dwarf novae (\cite{MW89}).

The quantitative and even qualitative behavior of the black
hole models depends on the prescription for $\alpha$.
In the case of a double valued, but radially constant
prescription, the outburst will tend to occur in the
inner disk, giving rise to somewhat slower rise, more
symmetric outbursts.  A prescription in which
$\alpha=\alpha_{0}(H/r)^{n}$ will give very small values
of $\alpha$ in quiescence where H/r is found to decrease inward.
This will yield a very long viscous time in the inner disk
and promote outbursts that begin in the outer disk
and propagate inward. This yields model outbursts with rapid rise
and slower decline, in accord with the observations for the
optical and soft X-ray light curves of the X-ray novae.

One of the interesting features of the black hole X-ray
novae is the tendency to show an exponential decline.
Simple models in which one quickly reduces the transfer rate
to a hot disk with with constant $\alpha$ generate geometrically
declining, not exponential, light curves.  Even models in
which the decline is driven by the cooling wave of the
disk instability tend to have geometrically declining light
curves with constant $\alpha$.  Mineshige et al. (1993) have argued that
to produce an exponential decline, the angular momentum of the
inner disk must be removed at a rate proportional to the
angular momentum.  They note that this tends to be the behavior
of disk instability models with $\alpha=\alpha_{0}(H/r)^{n}$
with n $\sim$ 1 - 2.  Cannizzo (1994) has also addressed this argument
by noting that both dwarf novae and the black hole transients
have exponential declines.  Cannizzo concluded that to
reproduce the exponential one needs $\alpha\propto r^{\epsilon}$,
with $\epsilon\sim~0.3 - 0.4$,
which is consistent with Mineshige et al.  Cannizzo carried
the argument one step further, however, by making the case that
the precise value of $\epsilon$ that leads to exponential
decline is itself a function of other parameters of the problem
such as the transfer rate and inner disk radius.  From this
he concluded that exponential decline requires some form of
feedback to operate in the disk to give just this behavior.
This may hint that the angular momentum transport process
is non-local, as the theories where internal waves play a
critical role imply.

These arguments have been extended significantly by Cannizzo,
Chen, and Livio (1995).  Cannizzo et al. used well-resolved
numerical studies to show that the width of the cooling front
can be approximated very closely by w = $\sqrt{Hr}$ and
that for such a behavior, exponential decay of the light
curve during the cooling wave phase is obtained only for
a prescription of the form $\alpha=\alpha_0(H/r)^n$ with
n very close to 3/2.  The critical point in their argument
is that angular momentum is removed from the hot part of the
disk by the advance of the cooling front and if the cooling
front velocity is proportional to $r$, as it is for $n\sim 1.5$,
then this loss of angular momentum is proportional to the
total angular momentum in the disk. This preferred value
of $n$ is consistent
with the theory of angular momentum transport by an internal wave-generated
dynamo driven by tidal instabilities at the outer edge of the disk
(\cite{VJD90}, \cite{VD92}).  Nevertheless, the
physical underpinnings of this behavior of the cooling wave were
not clear.  In particular, it is not clear why the cooling
front should have this width.  Given this width, it is possible
to argue that the cooling front should have the velocity characteristic
of torque induced mass flows with a radial length scale of $w$, i.e.
\begin{equation}
V_r\sim {\alpha c_s^2\over w\Omega}\sim \alpha c_s
\left({H\over r}\right)^{1/2}.
\label{eq:cwidth}
\end{equation}
Since $c_s$ near the cooling front is approximately constant, and
since $H\sim c_s/\Omega\propto r^{3/2}$, this gives a cooling front
velocity which is proportional to $r$ when $n=3/2$.
In what follows we will argue
that although this expression for the cooling front velocity is
approximately correct, the direction of causality has been reversed.
The cooling front width is a consequence of the cooling front velocity.

In this paper we present an analysis of the behavior of the
cooling wave and show that its propagation depends on the
viscous flow in the hot state and is nearly independent
of the actual cooling process and of the
state of the disk in the cool, quiescent material
that accumulates in the wake of the inward-propagating
cooling wave.  We argue, in agreement with Cannizzo, Chen,
and Livio, that the exponential decay gives strong evidence
for the presence of the cooling wave, and hence of the
disk instability phenomenon in general, and a powerful
constraint on the physical nature of the local viscosity.
The mechanisms that control the propagation of the cooling
front are discussed in \S 2.  Section 3 presents constraints
on the opacity and other functional forms of $\alpha$.
The relation of these results to the internal wave-driven dynamo
are presented in \S 4.  Summary and conclusions are given
in \S 5.

\section{The Cooling Front}

As the cooling front moves from large to small radii, it is preceded
by a rarefaction wave that lowers the temperature and column
density to the point where rapid cooling can set in.  A rough
sketch of the temperature and column density of the disk as a function
of radius is shown in figure 1, with the sudden steepening of the
temperature gradient to nearly a vertical line indicating the
onset of rapid cooling.  Figure 1 is merely illustrative of the cooling
process, but is consistent with the detailed figures given by CCL.
It is important to note that the radial distributions of
column density and temperature
of the disk are smooth power laws well inside the cooling front, but
that they drop below an extrapolation of these power laws before the
cooling front actually reaches them.  We will refer to the region
just inside of the cooling front
where departures from the power law distributions occur as the precursor
region.

We can understand the behavior of the cooling front by examining
the equations for the conservation of mass and angular momentum,
and the structure equations for a hot disk.
These are
\begin{equation}
\partial_t\Sigma=-{1\over r}\partial_r(r \Sigma V_r),
\label{eq:contin}
\end{equation}
and
\begin{equation}
V_r={2\over \Sigma\Omega r^2}\partial_r\left(r^3\alpha \Sigma
{c_s^2\over\Omega}
\partial_r\Omega\right),
\label{eq:torque}
\end{equation}
where $\Sigma$ is the gas column density, $\Omega(r)$ is the local rotational
frequency (proportional to $r^{-3/2}$ in a Keplerian disk), $\alpha$ is the
dimensionless viscosity, $c_s$ is the local sound speed, and $V_r$ is the
radial velocity.  The thermal structure of an optically thick disk is
determined
by its opacity source.  In general, the midplane temperature can be written as
\begin{equation}
T=B_1 \Sigma^{a_1} \alpha^{b_1} \Omega^{(2/3)c_1}.
\label{eq:thermal}
\end{equation}
In the hot portions of the disk for which the opacity can be
approximated by a Kramers law opacity,
$a_1=c_1=3/7$ and $b_1=1/7$.  Rapid cooling sets in for temperatures
below $T_{min}$, where
\begin{equation}
T_{min}\propto \alpha^{-1.1/7} \Omega^{-{3\over70}},
\label{eq:tmin}
\end{equation}
(cf. CCL, equation (4)).
It is important to note that $T_{min}$ has a very weak dependence on radius.

In the hot inner part of the disk, well away from the cooling front, radial
length scales are all comparable to $r$ and equation (\ref{eq:torque}) implies
a radial velocity of order
\begin{equation}
|V_r| \sim \alpha{c_s^2\over r\Omega},
\end{equation}
which is the usual result for a stationary accretion disk.  In fact,
the numerical simulations of CCL show that the inner disk is reasonably
well described by the standard stationary disk solution. The
mass transfer rate in the inner regions is changing with time, but is
fairly constant with radius.  Even though $\dot M$  reverses
sign as one approaches the cooling front, the column densities and
temperatures deviate appreciably from the stationary solution only close
to the cooling front.

Before proceeding with a discussion of the cooling front physics, it
will be useful to pause and consider the solutions to equations
(\ref{eq:contin}) and (\ref{eq:torque}) in the hot part of a disk.
Combining these expressions with equation (\ref{eq:thermal}), and
assuming that $\alpha\propto (H/r)^n$, we can show that
\begin{equation}
2\pi r \Sigma V_r={\pi\over r\Omega}\left(-{3\over2}{k_B\over\mu}B_1
\right)\partial_r\left(r^2\Sigma^{1+a_1}\alpha^{1+b_1}\Omega^{(2/3)c_1}
\right),
\end{equation}
or
\begin{equation}
\dot M=-C_2 r^{1/2}\partial_r(r^{\tilde q}\Sigma^{1/q}),
\label{eq:hstruct}
\end{equation}
where $C_2$ is a constant,
\begin{equation}
\tilde q\equiv 1+\left(1+{n\over2}\right)\left({1-c_1\over 1-{n\over2}b_1}
\right),
\end{equation}
and
\begin{equation}
q^{-1}\equiv 1+\left(1+{n\over2}\right)\left({a_1\over 1-{n\over2}b_1}
\right).
\label{eq:expq}
\end{equation}
If the disk is stationary then $\dot M$ will be a constant.  If the inner,
hot disk is slowly evolving, then $\dot M(t,r)\approx \dot M_0(t)$, and
integration over r gives:
\begin{equation}
\dot M_0\approx -{C_2\over 2} r^{\tilde q-{1\over2}}\Sigma_0(r)^{1/q},
\label{eq:m0t}
\end{equation}
or
\begin{equation}
\Sigma_0(r)\approx r^{({1\over2}-\tilde q)q}
\left({-2\dot M_0(t)\over C_2}\right)^q.
\end{equation}
It is useful at this point to define a function $f(r)\equiv \Sigma/\Sigma_0$.
If the hot, inner part of the disk is almost stationary then $f(r)\approx 1$
and $\partial_r f(r)\approx 0$ everywhere except close to the cooling front.
Then we have
\begin{equation}
\partial_t\Sigma=\partial_t(f\Sigma_0)\approx f\partial_t\Sigma_0=
{f\over q}\Sigma_0\partial_t \ln(\dot M_0(t)).
\label{eq:st}
\end{equation}
Meanwhile, from equations (\ref{eq:contin}), (\ref{eq:hstruct})
and (\ref{eq:m0t}) we have
\begin{equation}
\partial_t\Sigma=-{1\over 2\pi r}\partial_r \dot M
=-{\dot M_0(t)\over 2\pi r}\partial_r \left( f^{1/q}\left(1+{2\over q}
{\partial\ln f \over\partial \ln r}\right)\right).
\label{eq:fstruct}
\end{equation}
If we take $f\approx 1$ then this can be simplified and combined with
equation (\ref{eq:st}) to yield
\begin{equation}
{\Sigma_0q}\partial_t\ln\dot M_0\approx -{\dot M_0(t)\over 2\pi r}
({3\over q}\partial_r f+{2\over q}r\partial_r^2f),
\label{eq:st2}
\end{equation}
which has the solution
\begin{equation}
f\approx 1- \left({r\over r_e(t)}\right)^m
\end{equation}
where
\begin{equation}
m=2-q\tilde q+{q\over2}.
\end{equation}
Since $\Sigma$ becomes small as $r$ approaches $r_e(t)$ we can
interpret $r_e(t)$ as a measure of the outer edge of the hot part
of the disk.  More generally, the actual edge of the hot phase of
the disk will scale as $r_e(t)$, but will lie at some slightly smaller
radius.  Substituting this expression for $f(r)$ back into
equation (\ref{eq:st2}) we get
\begin{equation}
{\Sigma_0q}\partial_t \ln\dot M_0(t)={\dot M_0(t)\over 2\pi r^2}
{(2m^2+m)\over q}\left({r\over r_e(t)}\right)^m.
\end{equation}
At this level of approximation we conclude that if we consider any fixed $r$
well inside the radius of the cooling front we find that
\begin{equation}
r_e(t)\propto \left(-\partial_t\ln\dot M_0(t)\right)^{-1/m}
\left[-\dot M_0\right]^{{(1-q)\over m}}.
\end{equation}
Consequently, if $r_e(t)$ is an exponentially decreasing function of
time, then so is $\dot M_0(t)$.  Inasmuch as the bolometric luminosity
of the disk is determined by the mass accretion rate at small radii,
this implies a direct connection between an exponentially decreasing
radius for the hot portion of the disk, and an exponentially decreasing
bolometric disk luminosity.

One other point is worth mentioning here.  The radial velocity of the
gas is (cf. eq (\ref{eq:fstruct}))
\begin{equation}
V_r={\dot M\over 2\pi r\Sigma}={\dot M_0(t)\over 2\pi r\Sigma_0}
(f^{1/q-1}+{2\over q}rf^{1/q-2}\partial_r f)\approx
{\dot M_0(t)\over 2\pi r\Sigma_0}
f^{1/q-2}\left(1-(1+2{m\over q})\left({r\over r_e}\right)^m\right).
\end{equation}
This will have a zero when $r$ is a fraction of $r_e(t)$ $\approx$ 0.23
for a Kramers-law opacity and $n=1.5$.  At this point
$f\approx 0.8$, which is still close to unity.  The simulations of
CCL actually indicate that the radius of zero velocity will be at
about $0.38$ times the cooling front radius, but this is due to the
fact that $r_e$ is the radius at which the column density solution
goes to zero, rather than the radius at which it drops below the hot
phase minimum column density.  The point at which
V$_R$ = 0 is well defined in terms of the behavior of the velocity
even though the disk temperature and column density scarcely depart
from their steady-state values at that radius.  Beyond this radius,
the gas will move outward in the outer parts of the hot portion of the disk
as the disk material is uniformly stretched.
The column density will deviate from the stationary
solution only close to the cooling front.

We could explore this expansion further, but since it becomes rapidly
less accurate near the cooling transition, it seems unlikely to give
us means of deriving the cooling front velocity.  This does show that
the mass flow rate at small $r$ will decrease exponentially if
the radius of the hot portion of the disk decreases exponentially
{\it and if the cooling front moves slowly enough to allow the inner
part of the disk to stay close to a stationary solution.}  We will
see later that this condition is automatically satisfied.

In order to understand the cooling front velocity we need to focus
on the structure of the disk near the cooling front.
Let us consider some fiducial point just inside of the precursor region and
denote
quantities at that radius with a subscripted $p$.  Quantities evaluated at
the cooling front will be denoted with a subscripted $F$.  Mass
enters the precursor at a rate $\Sigma_p (V_r(r_p)+v_{cF})$, where $v_{cF}$
is the cooling front velocity.  This mass flow should be balanced, allowing
for some slight difference in $r_p$ and $r_F$ and the secular evolution of
the cooling front, by the flow of material into the region of rapid cooling.
At the onset of rapid cooling the gas
begins to cool at some large fraction of the thermal rate, which for an
optically thick disk is $\sim \alpha\Omega$.  If the radial
scale for temperature change is $L_F$ then from equation (\ref{eq:torque}) we
have
\begin{equation}
V_F \sim \alpha_F{c_F^2\over L_F\Omega_F},
\end{equation}
where $V_F$, the radial velocity at the cooling front,
is positive since the thermal gradient is strongly negative.
Inasmuch as the material is essentially freely cooling at this point
$L_F\sim V_F/(\alpha_F\Omega_F)$ which implies
\begin{equation}
V_F\sim\alpha_F c_F.
\end{equation}
Detailed comparison with CCL's results shows that the radial velocity at
the cooling front is actually only about $(1/6)\alpha_F c_F$.  On the
other hand, this fraction is constant in time and our main concern
here is with the scaling properties of the cooling front rather than
with a derivation of the various numerical factors which figure in the
actual solution.  Matter conservation implies that
\begin{equation}
\Sigma_p(V_r(r_p)+v_{cF})\sim \Sigma_F \alpha_F c_F.
\label{eq:cont}
\end{equation}
Here we have assumed that the mass velocity at the cooling front,
$V_F$, is much larger than the cooling front velocity, $v_{cF}$.

We can simplify equation (\ref{eq:cont}) further by arguing that $v_{cF}$
cannot be arbitrarily smaller than $V_r(r_p)$.  If it were, then
the disk would evolve ahead of the cooling front faster than the cooling
front could propagate.  In particular, the outer edge of the
hot portion of the disk, just
ahead of the cooling front, would become depleted of all matter before
the cooling front reached it.  Since equations (\ref{eq:thermal})
and (\ref{eq:tmin}) imply the existence of a minimal column density, below
which rapid cooling sets in, this situation is paradoxical.
A minimal speed for the cooling front is supplied by the rate at which
the outer edge of the hot portion of the disk would move inward
purely from the depletion of matter, i.e. the viscous accretion speed.
We conclude
that $v_{cF}$ is either scaling with $V_r(r_p)$ or becomes increasingly
larger than it as $r_F$ decreases.  Therefore
equation (\ref{eq:cont}) can be rewritten as
\begin{equation}
v_{cF}\sim {\Sigma_F\over \Sigma_p} \alpha_F c_F.
\label{eq:cont1}
\end{equation}

It is also evident that the cooling front cannot move arbitrarily faster than
$V_r(r_p)$.  If it did, then the inner disk would be unable to evolve
before the cooling front reached it.  Consequently, $\Sigma_p$
would be virtually unchanged from its value at the epoch when the
cooling front first formed at the outer edge of the disk.
{}From equation
(\ref{eq:torque}) and (\ref{eq:thermal}) it is straightforward to
show that for a Kramers law opacity and $\alpha$ constant
\begin{equation}
\Sigma\propto r^{-0.75},
\end{equation}
in a stationary disk.  Under the same assumptions, equations
(\ref{eq:thermal}) and (\ref{eq:tmin}) imply that $\Sigma_F$ can
increase almost as fast as $r$ (for a constant $\alpha$).  Equation
(\ref{eq:cont1}) then implies that $v_{cF}$ scales roughly as
$r^{(7/4)}\alpha_F$, or $r^{(7/4)+n/2}$.  This steep positive
scaling implies that $v_{cF}$ will drop rapidly as the cooling
front moves inward.  Substituting functional forms
of $\alpha$ that are consistent with disk observations changes the
values of the exponents slightly, but leads to the same qualitative
conclusion.  On the other hand, the condition that the mass flow
rate, $\dot M=rV_r\Sigma$, is a constant implies that $V_r$ {\it rises}
slowly as $r\rightarrow 0$.  We conclude that if the ratio of $v_{cF}$ to
$V_r(r_p)$ is large then it will decrease rapidly with
decreasing $r$.

Since neither a large nor small ratio of $v_{cF}$ to $V_r(r_p)$ is
sustainable we conclude that the cooling front will
evolve into a state where the two scale together, i.e.
\begin{equation}
v_{cF}\sim V_r(r_p)\approx \alpha_p{c_p^2\over r_p\Omega(r_p)}.
\label{eq:vel}
\end{equation}

We could have shortened the derivation of equation (\ref{eq:vel}) somewhat
if we had simply defined the fiducial point $p$ to lie at the radius
where the fluid velocity vanishes.  Then the mass flow into the precursor
region would have been $v_{cF}$ and $V_{r_p}=0$ by definition.  While
this would have eliminated most the argument preceding equation
(\ref{eq:cont1}), it would meant that the factors of
$(r_p/r_F)$ in our scaling relations could not be assumed to be
$\approx 1$.  In particular, we would have been forced to assume that
such factors were approximately constant, which is consistent with
the numerical simulations, but not otherwise justified in this paper.
It is more convenient to take $p$ close to the cooling front, although
with a radial scale length still of order $r$, and consequently a
radial velocity of the order given in equation (\ref{eq:vel}).

Equations (\ref{eq:thermal}), (\ref{eq:tmin}), (\ref{eq:cont1}), and
(\ref{eq:vel}) are all we need to solve for the scaling properties
of $v_{cF}$ for any given functional form of $\alpha$.
Combining equations (\ref{eq:cont1}) and (\ref{eq:vel}) we have
\begin{equation}
\alpha_p{c_p^2\over r_p\Omega(r_p)}\sim {\Sigma_F\over \Sigma_p} \alpha_F c_F.
\end{equation}
Since the precursor front is narrow we can replace $r_p$ with $r_F$.
(Actually, since we are only concerned with scaling laws, this argument
would work for a broad precursor as long as $r_p$ were some fixed fraction of
$r_F$.)  Rearranging terms we find
\begin{equation}
{\alpha_p\over\alpha_F}{T_p\Sigma_p\over T_F\Sigma_F}\sim {r_F\Omega_F\over
c_F}.
\label{eq:match1}
\end{equation}
In order to proceed beyond this point it is necessary to choose some form for
$\alpha$.  We start with the form $\alpha=\alpha_0(H/r)^n$, which
was used by CCL and is
motivated by the various theoretical and phenomenological arguments
cited in the introduction.
This implies $\alpha\propto (rT)^{n/2}$.  Consequently, equation
(\ref{eq:thermal}) becomes
\begin{equation}
T^{1-b_1{n\over2}}\propto \Sigma^{a_1} r^{b_1{n\over2}-c_1},
\label{eq:thermal2}
\end{equation}
and equation (\ref{eq:match1}) becomes
\begin{equation}
\left({T_p\over T_F}\right)^{1+{n\over2}}{\Sigma_p\over \Sigma_F}
\sim {r_F\Omega_F\over c_F}.
\label{eq:match2}
\end{equation}
Combining these two equations yields
\begin{equation}
{\Sigma_F\over \Sigma_p}\sim \left({c_F\over r_F\Omega_F}\right)^q,
\label{eq:match3}
\end{equation}
or,
\begin{equation}
v_{cF}\sim \alpha_F c_F \left({c_F\over r_F\Omega(r_F)}\right)^q
\sim c_F^{q+n+1}r_F^{{q+n}\over2},
\label{eq:match3a}
\end{equation}
where $q$ is defined in equation (\ref{eq:expq}).

CCL suggested $q=0.5$ on the basis of their numerical simulations.  Here
we see that its value is a function of the opacity law in the hot state
and the value of $n$.  For a Kramers law opacity we get $q=0.54$ for
$n=3/2$ with $q$ dropping to $0.5$ for $n=2$ and rising to $0.59$ for $n=1$.
In spite of its functional dependence on $n$ it is difficult to get
$q$ very different from $0.5$ for any reasonable choice of $n$.
The result is similarly insensitive to the exact opacity law.
If we consider electron scattering instead,
for which $a_1=2/3$, $b_1=1/3$, and $c_1=1/2$, we find that
$q=0.39$ for $n=3/2$.  Finally, we note that this
argument does not include the dynamics of the rapid cooling region,
or the cold state, at all.  These can be varied in any way that
preserves the existence of a rapid cooling zone without changing
the cooling front speed.

In their paper CCL proposed that $n$ should be close to $3/2$, since
that was the value that gave an acceptably exponential
decline in the disk luminosity as the cooling front propagated
inward.  More specifically $v_{cF}\propto r_F$ implies an exponential
decline, so treating $T_F$ (which is also $T_{min}$; eq (\ref{eq:tmin}))
and hence c$_F$
as approximately constant and $q=0.5$ implies $n=3/2$, as can be
seen from eq (\ref{eq:match3a}).  Strictly speaking, numerical models
show that $T_{min}$ depends somewhat on radius.
Given our form for $\alpha$ we can
rewrite equation (\ref{eq:tmin}) as
\begin{equation}
T_F\propto r^{({9\over140}-{1.1\over7}{n\over2})/(1+{1.1\over7}{n\over2})}.
\end{equation}
Substituting this into equation (\ref{eq:match3a}) we see that
\begin{equation}
v_{cF}\propto r^{0.949},
\label{eq:sca}
\end{equation}
for $n=1.5$.
and
\begin{equation}
v_{cF}\propto r,
\end{equation}
when $n=1.65$.
We see that our results suggest that slightly higher values of $n$ are
necessary to obtain a purely exponential decay when the temperature
at which cooling sets in depends somewhat on radius.  Of course, the
observations themselves suggest only approximately exponential decay.
The question here is whether or not the cooling front dynamics are
actually controlled by the preceding rarefaction wave as our model
assumes.  More recent
calculations (\cite{C96}) show that $n=1.625$ does indeed produce
a more nearly exponential form than $n=1.5$.

\section{Alternative Opacities and Functional Forms For $\alpha$}

We have seen that the link between the observed exponential decay of
soft X-ray luminosity of black hole binary systems and the conclusion
that $\alpha\propto (H/r)^{3/2}$ depends on the opacity law for the
hot state of the disk, as well as our initial assumption of a functional
form for $\alpha$.  In this section we will examine the consequences
of taking other opacity laws and ask whether or not there are other
forms for $\alpha$ that would do as well.  We will defer discussion
of a nonlocal model for $\alpha$, the internal wave driven dynamo,
until the next section.

It is helpful to begin by assuming that the hot disk is optically thick,
with an opacity law of the form
\begin{equation}
\kappa\propto \rho^A T^{-B}
\end{equation}
so that the disk opacity is
\begin{equation}
\tau\sim\kappa\Sigma\propto\Sigma^{A+1}T^{-B}H^{-A}\propto
\Sigma^{A+1}T^{-B-{A\over2}}\Omega^A.
\end{equation}
Invoking the equality between the energy generation rate per unit
area, $\dot M\Omega^2\sim \alpha\Sigma c_s^2\Omega$, and the
rate at which energy is radiated, $\sigma_B T^4/\tau$, we can
recover an equilibrium relationship of the form given in equation
(\ref{eq:thermal}) with
\begin{equation}
a_1={A+2\over 3+B+A/2},
\end{equation}
\begin{equation}
b_1={1\over 3+B+A/2},
\end{equation}
and
\begin{equation}
c_1=\left({3\over2}\right){A+1\over 3+B+A/2}.
\end{equation}

Limits on the plausible range of these parameters, and more importantly
for the exponent $q$ in equation (\ref{eq:expq}), can be deduced from
the requirement that the hot phase of the disk must be thermally and
viscously stable.  Thermal stability implies that for a fixed $\Sigma$
and $r$, a small positive deviation of $T$ away from equilibrium will
produce a larger increase in the cooling rate
than in the heating rate.  The cooling
rate per unit area is proportional to $T^4\tau^{-1}$ or
\begin{equation}
Q^-\propto T^4\tau^{-1}\propto T^{4+B+{A\over2}}.
\end{equation}
The heating rate per unit area is
\begin{equation}
Q^+\propto \dot M\propto \alpha T.
\end{equation}
If we use
\begin{equation}
\epsilon\equiv {\partial \ln \alpha\over\partial \ln T},
\end{equation}
to parameterize the dependence of $\alpha$ on temperature, then the
hot phase will be thermally stable if
\begin{equation}
1+\epsilon<4+B+{A\over2}.
\end{equation}
Remembering that for $\alpha\propto(H/r)^n$ we have $\epsilon=n/2$, we
see that equation (\ref{eq:expq}) implies that $q\rightarrow 0$
as the hot phase approaches the threshold of thermal instability.
In other words, as we consider opacities for the hot phase that
bring it closer and closer to a loss of thermal stability, the cooling
front moves closer and closer to its maximum speed of $\alpha_F c_F$.

The condition that the hot phase is viscously unstable is that
\begin{equation}
{\partial\dot M\over\partial\Sigma}<0,
\end{equation}
where thermal stability is assumed.  This can be rewritten as
\begin{equation}
1+{\partial\ln\alpha\over\partial\ln\Sigma}+{\partial\ln T\over\partial
\ln\Sigma}>0,
\end{equation}
or
\begin{equation}
{\partial\ln T\over\partial \ln\Sigma}\left(1+{\partial\ln\alpha
\over\partial\ln T}\right)>-1.
\end{equation}
We can see by comparing this result to equations (\ref{eq:thermal})
and (\ref{eq:expq}) that viscous instability will set in only
when $q$ passes through $\infty$ to negative numbers.  In other
words, an arbitrarily slow cooling front can be produced by
letting the hot phase go to the threshold of viscous instability.

It seems odd that it is possible to get a cooling front moving more
slowly than the viscous accretion speed.  The reason this works is
that when $a$ (as defined in equation (\ref{eq:thermal})) is negative,
the rarefaction wave that precedes the cooling front actually raises
the temperature of the gas, making it more difficult for it to reach
the minimum hot phase temperature.  In fact, it is difficult to see
how we can get a cooling front when this is the case.
If we adopt the more reasonable requirement that
$T$ increase with $\Sigma$ (i.e. $a>0$) then the slowest possible cooling
front speed is just $\alpha_F c_F (H_F/r_F)$, when $a=0$ (and
$q=1$).  In other words, as the midplane temperature becomes
insensitive to the column density, the cooling front speed converges
to the accretion velocity in the hot state.  It seems that for any
reasonable opacity law for the hot state, the exponent $q$ will fall
in the interval $[0,1]$.  Since the limits describe fairly extreme
situations, we expect that typically $q$ will fall somewhere near the
middle of this interval, even when the hot state does not have
a Kramer's law opacity.

Now we consider the effect of using a different functional form for
$\alpha$.  The behavior of the cooling front depends only on the properties of
the hot state, including the functional form of $\alpha$.  This implies
that a model where $\alpha$ is given by a pair of values, i.e.
$\alpha=[\alpha_{hot},\alpha_{cold}]$ is equivalent to taking $n=0$
in the preceding section.  In this case the cooling front speed is
\begin{equation}
v_{cF}\sim \alpha_{hot}c_F\left({c_F\over r_F\Omega(r_F)}\right)^{1\over 1+a}
\propto c_F^{1.7} r^{0.35},
\end{equation}
which is unacceptably different from $v_{cF}\propto r$.

Alternatively, one could take $\alpha\propto r^k$.  This form has the
problem that $\alpha$ does not increase as one passes from the cold
state to the hot state, and for that reason is unlikely to produce
a fit to the entire outburst cycle.  However, we can ask the narrower
question of whether or not such a form could produce an exponential
decay.  Since $\alpha$ is a function of $r$ only, and since $r_p$
and $r_F$ differ by only a constant factor (which will be close to one
in most cases) equation (\ref{eq:thermal}) implies that
\begin{equation}
{T_p\over T_F}=\left({\Sigma_p\over\Sigma_F}\right)^{3/7}.
\end{equation}
Then equation (\ref{eq:match1}) implies
\begin{equation}
\left({\Sigma_p\over \Sigma_F}\right)^{10/7}\sim {r_F\Omega_F\over c_F}.
\label{eq:match4}
\end{equation}
Combining this result with equations (\ref{eq:cont1}) and
(\ref{eq:tmin}) we find that
\begin{equation}
v_{cF}\sim \alpha(r_F) c_F\left({c_F\over r_F\Omega(r_F)}\right)^{0.7}
\propto r^{0.405+0.866k}.
\end{equation}
We conclude that for $\alpha\propto r^k$
an exponential decay law will follow from $k\approx 0.69$.

\section{The Internal Wave Driven Dynamo Model}

The internal wave driven model (\cite{VJD90}, \cite{VD92}, and \cite{VD93})
predicts that $\alpha$ has the
form assumed in CCL, with $n\approx 3/2$.  (The precise value
depends somewhat on the way the wave cascade is modeled, with
$n$ as low as $4/3$ possible if every level in the cascade contributes
maximally to the dynamo process.)  While it is exciting to
see this prediction validated by CCL,  the internal wave driven
dynamo model is inherently nonlocal, and it is unclear whether or
not the arguments given in the preceding section can be applied to it.
More specifically, in this model the local dynamo activity, and the
consequent turbulent viscosity, depend on the amplitude of the
internal wave field.  These waves propagate inward and are excited
by tidal instabilities near the outer edge of the disk (\cite{G93},
\cite{RG94}, \cite{RGV96}, \cite{VZ96}).
Normally their amplitude can be estimated by balancing nonlinear
dissipation with their linear amplification and focusing as
they propagate inward.  This implies a mean square wave amplitude
proportional to $(H/r)$.  However, near a cooling front the
waves travel across a region of relatively sharp increase in this ratio
and one would expect that the wave amplitudes would fall well below their
usual saturation value. It follows that in order to test the
consistency of the internal wave driven dynamo model there
are several issues that must be addressed.  First, what is
the effect of decreasing $\alpha_F$ below the value inferred
from the equilibrium expression?  Second, how does the amplitude
of the internal waves change as they cross the cooling front?
Third, how does this decrease translate into a reduced value for
$\alpha_F$?

We could add to these the question of whether or not the amplitude
of the $\alpha$ inferred from the luminosity decay in X-ray binaries
is consistent with this model. Unfortunately, the model is not
yet sophisticated enough to predict this constant, so this test must
be deferred until a later time.

We will assume from the start that away from the cooling front
$\alpha=\alpha_0(H/r)^n$ and the hot state opacity is given by
Kramers law.  Then we define the softening factor $f$ by
\begin{equation}
\alpha_F=\alpha_0 f\left({H\over r}\right)^n.
\end{equation}
In this case equation (\ref{eq:match1}) implies
\begin{equation}
{r_F\Omega_F\over c_F}\sim {\alpha_p\over\alpha_F}{T_p\Sigma_p\over
T_F\Sigma_F}
=f^{-1} \left({T_p\over T_F}\right)^{1+n/2}{\Sigma_p\over\Sigma_F}.
\label{eq:match5}
\end{equation}
{}From equation (\ref{eq:thermal}) we have
\begin{equation}
\left({T_p\over T_F}\right)^{1-{n\over14}}=f^{-{1\over7}}\left(
{\Sigma_p\over\Sigma_F}\right)^{3/7},
\end{equation}
for a Kramer's law opacity.
Combining this with equation (\ref{eq:match5}) we get
\begin{equation}
{r_F\Omega_F\over c_F}\sim f^{-1-{1+n/2\over 7-n/2}}
\left({\Sigma_p\over\Sigma_F}\right)^{1/q},
\label{eq:srat}
\end{equation}
where $q$ is given in equation (\ref{eq:expq}).  This gives us
the dependence of the ratio of column densities on $f$.  The
cooling front velocity is also modified by the rise in $T_{min}$
(and therefore $c_F$) caused by the drop in $\alpha_F$.  Combining
equations (\ref{eq:tmin}), (\ref{eq:cont1})
and (\ref{eq:srat}) we conclude that
\begin{equation}
v_{cF}=v_{cF0}f^k,
\label{eq:vk}
\end{equation}
where
\begin{equation}
k={9.3\left(1+{n\over2}\right)\over (10+n)(14+1.1n)},
\label{eq:k}
\end{equation}
and $v_{cF0}$ is the cooling front velocity when $f=1$.
The exponent $k$ is small and relatively insensitive to $n$.  For
$n$ close to $1.5$, $k\approx 0.09$.  It seems clear that
$f$ has to vary strongly with radius in order to change the
scaling of the cooling front speed with radius.

A more significant problem is that for $f$ very small, the characteristic
S-shape equilibrium curve in the $\Sigma-T$ plane disappears, the
cooling transition becomes gradual, and the front speed dynamics become
more complicated.  When this happens will depend
critically on the equilibrium curve for the cold state, which has
not previous affected our calculations.  We will not discuss this question
further, but note that it should be a focus of subsequent work.

In order to decide whether or not the small change in the scaling
relationship induced by $f$ will create problems for the internal
wave driven dynamo model, we need to examine the scaling of $\alpha$
in this model with the wave amplitude.  For a review of this model
see Vishniac \& Diamond (1993).  We will quote selected results here.
The internal waves responsible for driving the dynamo consist of
slightly non-axisymmetric ($m=1$) waves whose amplitude is determined
by the balance between linear amplification, which occurs at a rate
of roughly
\begin{equation}
\tau_{amp}^{-1}\sim {V_{group}m\over r}\sim \left({H\over r}\right)\Omega,
\end{equation}
and nonlinear damping, which occurs at a rate of roughly
\begin{equation}
\tau_{nonlinear}^{-1}\sim {\cal M}^2\left({\Omega^2\over\bar\omega}\right).
\end{equation}
Here $\bar\omega$ is the comoving frequency of the wave, $V_{group}$ is
the radial group velocity of the waves,  and $\cal M$ is their
Mach number.  For these waves
\begin{equation}
V_{group}\sim\left({\bar\omega\over\Omega}\right)^2c_s.
\end{equation}
Nonlinear interactions will keep $\bar\omega/\Omega$
of order unity as long as the underlying disk changes only on length scales
of order $r$.  This implies that normally ${\cal M}^2\sim (H/r)$.  However,
as the waves pass through the cooling front their group velocity increases.
Conservation of wave energy flux implies
\begin{equation}
\left({{\cal M}_F\over{\cal M}_{cold}}\right)^2= {\Sigma_{cold}\over\Sigma_F}
\left({T_{cold}\over T_F}\right)^{3/2}.
\end{equation}
We will define this ratio as the factor $D$.  Since the waves start out
in the cold state with an amplitude that roughly balances linear amplification
with nonlinear dissipation, and since the nonlinear dissipation rate is
proportional to ${\cal M}^2$, $D$ is also the ratio of nonlinear dissipation
rate to the linear amplification rate just inside the cooling front.
In CCL the ratio of column
densities was of order $10$, while the temperature ratio was of order $10^2$,
implying $D\sim 10^{-2}$.  However, this temperature ratio results from
extrapolating using a power-law fit to the cold state models with midplane
temperatures of a few thousand degrees.  While the cold state opacities
are not completely known, the actual midplane temperatures for the cold
state probably fall in the range $1-3\times 10^3$ K.  This implies a
temperature
ratio of somewhat more than $10$ across the cooling front.  On the other hand,
the column density ratio should be of order $\Sigma_p\over\Sigma_F$, since
the material moves rapidly only near the cooling front itself and will
have a velocity, relative to the cooling front, of order $v_{cF}$ both in
the the cold state just outside of the cooling front and just inside the
precursor region.  Consequently,
this ratio will be largely unaffected by uncertainties in the cold state
$\Sigma-T$ relation.  It seems safe to conclude from this that $D$ is
generally of order $10^{-1}$.

How can we obtain $f$ from $D$?  This turns out to hinge on which waves
are responsible for driving the dynamo process.  For an `$\alpha$' driven by
magnetic
turbulence the value of $\alpha$ is roughly the dynamo growth rate divided
by $\Omega$.  For the wave-driven dynamo this is
\begin{equation}
\alpha\sim{\Gamma_{dynamo}\over\Omega}\sim\left({\cal M}^2 {m\over\bar\omega
\tau_{nonlinear}}\left({H\over r}\right) \Delta\right)^{1/2}.
\label{eq:wdyn}
\end{equation}
Here $\Delta$ is the fractional asymmetry of the wave field.  Only modes with
the same sign of $mk_r$ add coherently.  If we restrict ourselves to
the $m=1$ modes  (cf. Vishniac, Jin \& Diamond 1990) then this implies
$\alpha=\alpha_0(H/r)^{3/2}$, where the constant $\alpha_0$ is undetermined.
However, the process by which the internal waves dissipate is through a
period doubling cascade, which is eventually truncated through interactions
with the magnetically driven turbulence.  The dissipation rate of the
waves by the turbulence goes as $\alpha (\Omega^3/\bar\omega^2)$ and the
end point of the cascade is obtained by equating this to
$\tau_{nonlinear}^{-1}$.
If we assume that $\Delta\approx 1$ throughout the cascade (cf.
Vishniac \& Diamond 1992) then this implies $\alpha=\alpha_0(H/r)^{4/3}$.

The difference of a factor of $(H/r)^{1/6}$ has not previously been important,
but it may be in this context.  In addition, for the purpose of determining
$f(D)$ it matters a great deal whether or not the dynamo process is controlled
by the waves at the top or the bottom of the cascade.  Conservation of
radial and azimuthal momentum in the cascade turns out to imply that
$\Delta$ has to decrease at least as fast as $\bar\omega/\Omega$ as
one goes to smaller $\bar\omega$, which rules out $\alpha\propto (H/r)^{4/3}$.
However there are limits to how quickly $\Delta$ can decrease with
$\bar\omega$.
{}For any wave $\bar\omega$ is a function of position,
and an asymmetry in the wave field is created by normal wave propagation.
The size of this effect is roughly
\begin{equation}
\Delta\sim V_{group}\tau_{nonlinear}\partial_r\ln\bar\omega\sim
\bar\omega\tau_{nonlinear}{mH\over r}.
\end{equation}
Combining this result with equation (\ref{eq:wdyn}) gives
\begin{equation}
\alpha\sim{H\over r}{\cal M}|m|.
\end{equation}
Conservation of wave energy in the cascade implies that
${\cal M}^2\propto\bar\omega^{1/2}$.  If $m$ increases as we go down
the cascade though a random walk process then $m\propto\bar\omega^{-1/2}$
and invoking the condition for the turbulent truncation of the cascade
gives
\begin{equation}
\alpha=\alpha_0\left({H\over r}\right)^{10/7},
\label{eq:aw}
\end{equation}
when $D=1$.  In other words, although the cascade does dominate the
total helicity, and the subsequent value of $\alpha$, it increases the
dynamo growth rate by only a small factor.
When $D<1$ the cascade will extend to lower $\bar\omega$.  In addition,
the upper end of the cascade will change slightly, since for small
$D$ the linear evolution of the waves is much more important than nonlinear
dissipation.  For simplicity we will ignore such effects here.  Ignoring
the entire cascade and considering only the fundamental modes gives
$\alpha\propto D$.  Attempts to model the effect of $D<1$ on the whole
cascade change this very slightly.  This implies that the internal
wave driven dynamo contribution to $\alpha$ drops by more than an
order of magnitude across the cooling front.  The exact amount and
its scaling with radius will depend on details of the cold state
opacity.

{}Fortunately, there is a simpler solution to this problem.  The turbulence
induced by magnetic field instabilities can support a dynamo in a shearing
environment even in the absence of any long term average helicity
(\cite{VB96}).  For an accretion disk this gives $\alpha\propto (H/r)^2$.
The constant of proportionality is different from $\alpha_0$ and is
again unknown.
Lacking an accurate estimate we will assume that it is
comparable to $\alpha_0$.  Since $(H/r)^{1/2}$ is only
slightly less than $10^{-1}$ for these disks, it seems reasonable to
take $f\propto (H/r)^{1/2}$.  From equations (\ref{eq:vk}), (\ref{eq:k}),
and (\ref{eq:aw}) we see that this implies the cooling wave
moves at the speed one would expect for $n=10/7+0.045\approx 1.47$.

Of course, if $f$ is small enough the dynamics of the cooling front
may be significantly altered.  Further work on this, including a
derivation of the scaling coefficients of the internal wave driven
dynamo and the incoherent dynamo, are necessary to fully answer
the question of whether or not the internal wave driven dynamo
is consistent with observations of post-outburst luminosity
decline in accretion disks.

\section{Summary and Conclusions}

We have constructed a simple model for the propagation of cooling
fronts in accretion disks which reproduces the numerical results of
Cannizzo, Chen, \& Livio (1995).  In this model the cooling front speed
is determined by the rarefaction wave that lowers the disk temperature
to the point where rapid cooling can set in.  We find that the speed of the
cooling front scales as
\begin{equation}
v_{cF}\sim \alpha_F c_F \left({c_F\over r_F\Omega(r_F)}\right)^q,
\end{equation}
where the subscript $F$ refers to the radius where the disk falls out
of thermal equilibrium and begins rapid cooling.  The coefficient
$q$ is given in equation (\ref{eq:expq}) and depends on both the opacity
law and the functional form of $\alpha$.  However, $q$ will be
close to $1/2$ for most models of the disk hot state.  Somewhat
surprisingly, a local reduction of $\alpha$ near the cooling front
has only a modest effect on this result.

One striking aspect of our derivation is that we do not appeal to
any aspect of the structure of the disk at radii greater than $r_F$,
which marks the onset of rapid cooling.  This stands in contrast
to the suggestion by CCL that the cooling front velocity is determined
by its width, measured from the onset of rapid cooling to its finish.
We have not discussed the structure of this region here, but we note
that given a cooling front velocity determined by the structure of
the disk inward from the cooling front, the width of the cooling front
itself can be estimated by inverting equation (\ref{eq:cwidth}).
In other words, the scaling of the cooling front width is a consequence
of the speed of the cooling front, not its cause.

Our success in modeling the propagation of cooling fronts as
rarefaction waves suggests a similar effort could be made to
model heating fronts as compressional waves.  We have not yet
done this, but expect to examine this problem in a future paper.

Aside from the internal wave driven dynamo model,
we have not discussed models for $\alpha$ which are consistent
with the results of this paper.  While this is largely from a lack of
suitable candidates, there is one other prediction of $\alpha$
with the required form (\cite{MM83}).  However,
this estimate is based on using large scale buoyant cells
driven by magnetic buoyancy via the Parker instability
(cf. \cite{P75}).  Zweibel \& Kulsrud (1975)
showed that sufficiently strong turbulence in a shearing
environment would suppress the Parker instability.
Vishniac \& Diamond (1992) pointed out that the Balbus-Hawley instability
(\cite{V59}, \cite{C61}, \cite{BH91}) always leads to a level of turbulence
which is sufficiently strong by this criterion.  In fact, the linearly
unstable modes of an azimuthal magnetic field suffer turbulent mixing
at a rate roughly equal to the local orbital frequency. We conclude
that the magnetic buoyancy driven model is not consistent, in its
original form, with the dynamics of magnetic fields in accretion disks.

There are several conclusions to be drawn from this work.

{}First, we have provided a simple analytic derivation which supports
the conclusion of CCL that the exponential decay of the luminosity of
black hole disk systems following outbursts is consistent with a local law
for the dimensionless disk viscosity $\alpha\propto (H/r)^n$ if,
and only if, $n$ is approximately $3/2$.

Second, given this scaling for $\alpha$ we find that disk systems
in general should exhibit approximately exponential luminosity
decay from peak luminosity whenever the hot state opacity follows a
simple power law.  Exceptions will involve hot state $(\Sigma, T)$
relations which are either unusually close to thermal instability,
in which case the cooling front velocity can approach $\alpha_F c_F$,
or in which $T$ is extremely insensitive to $\Sigma$, in which case
the cooling front velocity will approach the accretion velocity in
the inner disk.

Third, this result implies that any $\alpha$ scaling for which
$\alpha$ is constant in the hot state is in conflict with current
observations.  This includes models in which $\alpha$ is given
by $[\alpha_{hot},\alpha_{cold}]$, where $\alpha_{hot}$ is a
constant.

{}Fourth, since the cooling front speed depends only on the hot state,
other models for $\alpha$ can also give exponential decays, although
they may fail on other grounds.  For example, if $\alpha\propto r^{2/3}$,
then we can obtain a roughly exponential luminosity decay in spite of the
fact that this law is insensitive to the local temperature.

{}Fifth, this result is apparently compatible with the internal wave
driven dynamo model for disk viscosity.  This does not follow
trivially from the prediction that $\alpha\propto (H/r)^n$,
where $n$ is approximately $3/2$ in a stationary disk.
The waves reach the cooling front after traveling through the cold part of the
disk.  Consequently, the $\alpha_F$ induced by the internal wave driven
dynamo is greatly reduced.  Here we have relied on an independent
mechanism, the incoherent dynamo, to give a minimal value for $\alpha_F$.
The final scaling law obtained in this way lies within observational
limits.  Clearly further work on these dynamo mechanisms, and on
the nature of the cold disk state, would be helpful for providing
a definitive answer to this question.  Still,  this
is the only internally consistent model for $\alpha$ which is
constructed from first principles and which satisfies the cooling wave
constraint.  We note that CCL have shown that the value of
$\alpha_0$ can be estimated from the luminosity decay rate.  This
value has not been calculated for the internal wave driven dynamo
model, but when it is the existence of an observationally motivated
estimate will provide another critical test of the model.

{}Finally we note that this whole analysis is predicated on the
assumption that a cooling wave exists in the decline of the
light curve of transient black hole candidates and related
systems.  While the evidence is indirect, one can thus regard
the exponential decline as a strong argument that a cooling wave
is the fundamental mechanism of the decline of these
transients. Furthermore, it adds to the evidence that
the accretion disk ionization instability is the
underlying physical cause of the transient outburst phenomenon.

\acknowledgements
It is a pleasure to acknowledge several helpful conversations with
John Cannizzo, whose calculations inspired this paper.  This work
has supported in part by NASA through grant NAG5-2773 to ETV and
through grants NAGW-2975 and NAG5-3079 to JCW.

\clearpage
\begin{figure}
\plotone{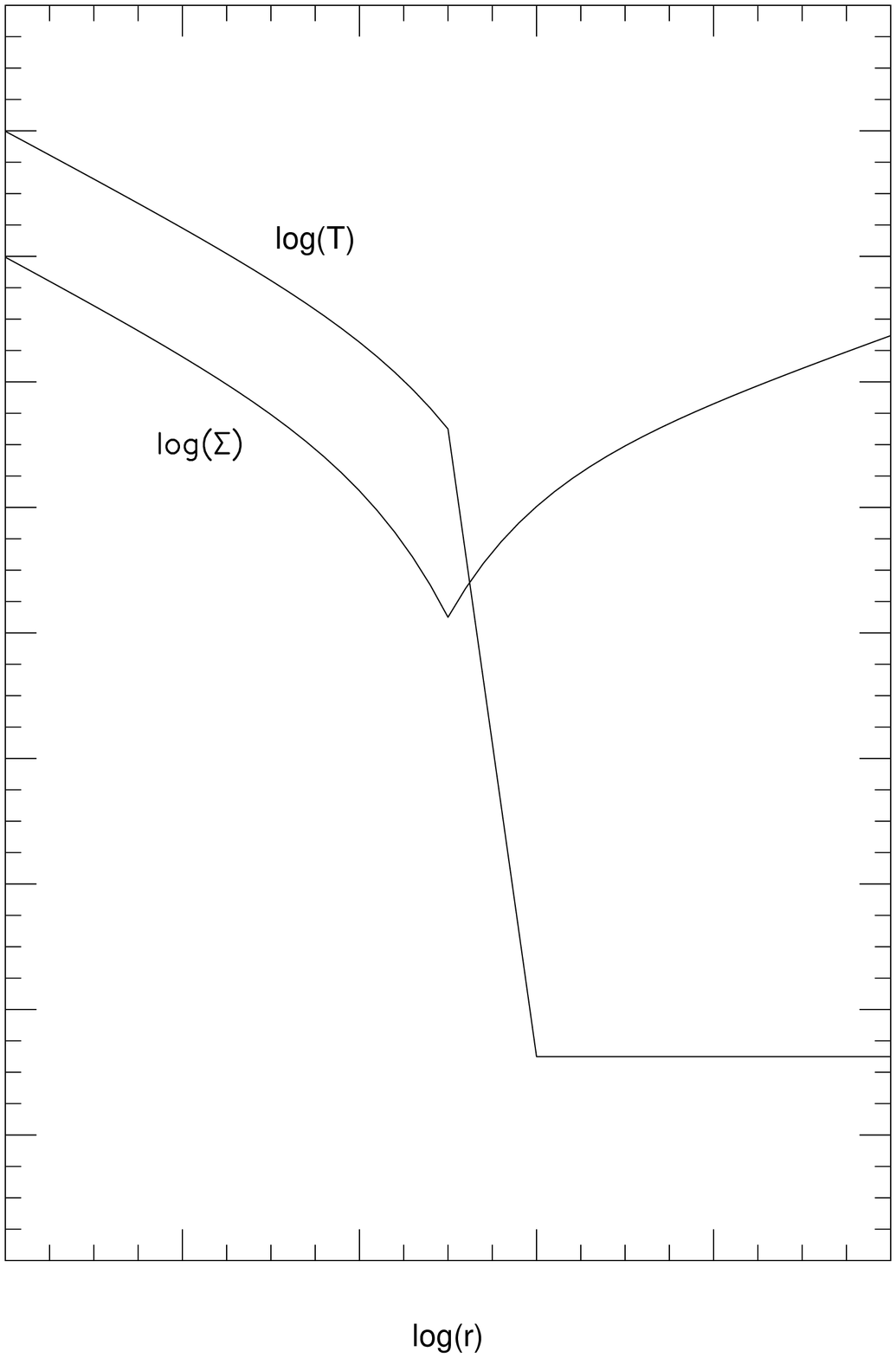}
\caption{A schematic of the variation of temperature $T$ and column
density $\Sigma$ as a function of radius $r$ across a cooling front
propagating towards small $r$.}
\end{figure}
\end{document}